\documentclass{article}

\begin{document}

\title
{How Quantum Entanglement Helps to Coordinate Non-Communicating
Players}

\author{Gleb V. Klimovitch \\
Stanford University \\
Stanford, CA\\}
\date{May 10, 2004}
\maketitle
\begin{abstract}
We consider a coalitional game with the same payoff for all
players. To maximize the payoff, the players need to use one
collective strategy, if all players are in certain states, and the
other strategy otherwise. The current state of each player changes
according to external conditions and is not known to the other
players. In one example of such a game, quantum entanglement
between players results in the optimal payoff thrice the maximal
payoff for unentangled players.
\end{abstract}

\section{Introduction}
Suppose that several players with the same payoff in the game need
to coordinate their moves, even though they cannot exchange
information with each other during the game. The players can agree
in advance on their collective strategy. However, the optimal
strategy for each player may depend on what the other players
"see" during the game, the knowledge the players cannot
communicate to each other.

We concentrate on a specific example how to coordinate
non-communicating players. In our example, the "quantum" solution,
in which players share entangled qubits pairs, is clearly superior
to the classical one.

\section{Discussion}

Consider a two-player game where players somehow need to switch
between the two opposite strategies. If {\em both} players are in
a certain state, they score by making the same simultaneous moves.
Otherwise, i.e. when none or only one player is in this state, the
players score my making different moves at the same time. Without
loss of generality, let each player have a binary statespace
$\{0,1\}$ and make binary moves $\{A,B\}$. Let $q_{ij}$ be the
state-dependent probabilities to make different simultaneous moves
when players one and two are in states $i$ and $j$, respectively.
(The arbiter cannot observe the {\em exact} "probabilities".
Instead, we could use {\em empirical} probabilities and let the
number of moves go to infinity, but such mathematical strictness
hardly adds any physical insight). The payoff $P$ is defined as:
\begin{equation}
P=\frac{q_{00}}{\max\{q_{01}, q_{10}, q_{11}\}}\;.
\end{equation}

Let us compare the optimal strategies and payoffs in classical and
quantum case.

\subsection{Classical strategy}
We can start with an arbitrary binary sequence $X_0$ of
length$N\gg 1$. For easier comparison with quantum case, let $X_0$
be a coin-flipping sequence: \newline $X_0 \sim Bern(1/2)$. Let us
choose nonzero probability $q<1/3$. Sequence $X_1$ is generated
from $X_0$ by "flipping" $qN$ bits of $X_0$. Sequence $X_2$ is
generated from $X_1$ by flipping $qN$ "other" bits of $X_1$.
Finally, sequence $X_3$ is generated from $X_2$ by flipping $qN$
bits of $X_2$, which have not changed  when generating $X_1$ and
$X_2$. The Hamming distances $d$ between the sequences are then
given by:
\begin{equation}
d(X_{\alpha}, X_{\beta})=qN|\alpha-\beta|\;;\; \alpha,\beta
=0,...,3\;.
\end{equation}

Player one transmits sequences $X_0$ and $X_2$ in states "$0$" and
"$1$", respectively. Player two sends sequences $X_3$ and $X_1$ in
states "$0$" and "$1$", respectively. Thus the state-dependent
probabilities to make different moves $q_{ij}$ are given by:
\begin{equation}
q_{01}=q_{10}=q_{11}=q\;,\; q_{00}=3q\;.
\end{equation}

The optimal classical payoff is then $P=3$. (The optimality check
is a simple exercise.) The same payoff occurs in the limit $q
\rightarrow 0$, if each subsequent sequence $X_{\alpha+1}$ is
obtained by passing its preceding sequence $X_{\alpha}$ through a
binary symmetric channel with crossover probability $q$.

\subsection{Quantum strategy}
The simultaneous moves depend on the spin measurements of qubit
singlets shared between the players. The players agree on the four
measurement directions, say in $XY$ plane, defined by angles
\begin{equation}
\phi_{\alpha}=\alpha\,\delta\;;\;\alpha=0,...,3\;;\;\delta\ll 1
\end{equation}

Player one measures the component of qubit spin along $\phi_0$ and
$\phi_2$ in states "$0$" and "$1$", respectively. Player two
measures the component of qubit spin along $\pi+\phi_3$ and
$\pi+\phi_1$ in states "$0$" and "$1$", respectively. Then the
angle between the measurement directions for the players is
$(\pi-3\delta)$, if both players are in state "$0$", and
$(\pi\pm\delta)$ otherwise. Each player makes "A" ("B") move when
the spin projection on the measurement direction is positive
(negative), respectively.

The state-dependent probabilities to make different moves $q_{ij}$
are given by \cite{Bell}:
\begin{equation}
\nonumber
q_{00}=\frac{1-\cos3\delta}{2}=\frac{9}{4}\delta^2+O(\delta^4)\;,
\end{equation}
\begin{equation}
\nonumber
q_{01}=q_{10}=q_{11}=\frac{1-\cos\delta}{2}=\frac{1}{4}\delta^2+O(\delta^4)\;.
\end{equation}

The optimal classical payoff is then $P \rightarrow 9$, achieved
in the limit $\delta \rightarrow 0$. (The proof of optimality is
to be verified and reported elsewhere).

\subsection{Generalization}
It is straightforward to calculate the optimal strategies and
payoffs under more general conditions, by means of inequalities
like
$$
q_{00} \le (\sqrt{q_{01}}+\sqrt{q_{10}}+\sqrt{q_{11}})^2
$$
for quantum case and
$$
q_{00} \le q_{01}+q_{10}+q_{11}
$$
for classical case. Other generalizations will be reported
elsewhere.

\section{Conclusions}
We considered a game-theoretical example, in which the payoff
depends on the correlation between simultaneous moves of
non-communicating players. Player coordination by quantum
entanglement triples the optimal payoff.

Similar techniques apply to other games with restrictions on
interplayer communication.

\end{document}